\documentclass[preprint,5p,twocolumn,10pt]{elsarticle}
\usepackage{amsmath,amssymb,bm,booktabs,array}
\usepackage{newtxtext}
\usepackage[varg]{newtxmath}
\usepackage{graphicx}
\usepackage[font=small,labelfont=bf,labelsep=period,justification=justified,singlelinecheck=true]{caption}
\usepackage{microtype}
\usepackage{url}
\usepackage[colorlinks=true,linkcolor={blue!50!black},citecolor={blue!50!black},urlcolor={blue!50!black}]{hyperref}
\graphicspath{{figures/}}
\biboptions{sort&compress}
\journal{Electric Power Systems Research}
\begin{document}
\begin{frontmatter}
\title{Passive Identification of Unequal Phase Grid Impedances by Phase-Domain Source Cancellation}

\author[hw]{Songyang Zhang\corref{cor1}}
\ead{so10@ualberta.ca}
\author[hw]{Xun Gong\corref{cor1}}
\ead{xun.gong@huawei.com}
\author[hw]{Bo Cao}
\ead{bo.cao@huawei.com}
\author[ut]{Reza Iravani}
\ead{reza.iravani@utoronto.ca}
\cortext[cor1]{Corresponding authors.}

\affiliation[hw]{organization={Montreal Research Centre, Huawei Technologies Canada Co., Ltd.},
  addressline={7101 Avenue du Parc}, city={Montreal}, state={QC}, postcode={H3N 1X9}, country={Canada}}
\affiliation[ut]{organization={The Edward S. Rogers Sr. Department of Electrical and Computer Engineering, University of Toronto},
  addressline={10 King's College Road}, city={Toronto}, state={ON}, postcode={M5S 3G4}, country={Canada}}

\begin{abstract}
Passive grid impedance estimators that return a single resistance--inductance pair cannot represent a grid whose phase impedances are unequal. This paper identifies a separate pair for each phase from recorded point-of-common-coupling (PCC) voltage and current without injected perturbations. In each phase, the unknown grid-source voltage is cancelled by a periodic difference (PD) or by a second-order-generalized-integrator notch difference (SOGID), and moving-window integration (MWI) replaces the current derivative by an integral and an endpoint feature. The resulting two-parameter regression requires a period or notch-frequency estimate but no phase-locked-loop angle. A voltage-balance analysis gives the source residual left by frequency mismatch and bounds the parameter error by the ratio of this residual to the conditioning of the two features. In the unequal-phase interval (8.3--9.8~s) of a 10-s EMTP grid-forming record, SOGID-MWI attains phase-averaged R/L errors of 5.09/0.33\%; a shared value has a minimum error of 41.67\% for the phase ratio 1:0.5:2. On a 50-s constant-impedance grid-following record, the errors are 0.56/0.06\% over the 20--30-s synchronization disturbance. In a controlled 40-dB noise study, a noise-referenced update rule reduces the white-noise GFL error from 55.9/65.0\% to 12.3/11.5\%.
\end{abstract}
\gdef\useauthors{Songyang Zhang, Xun Gong, Bo Cao, Reza Iravani}
\hypersetup{pdfkeywords={Grid impedance estimation, passive identification, phase-domain identification, unbalanced grid, recursive least squares, grid-forming converter, weak grid}}
\end{frontmatter}

\section{Introduction}
Knowledge of the grid impedance seen from the point of common coupling (PCC) supports protection, control and operation of converter-rich networks. Representative uses include hosting-capacity monitoring, adaptive protection setting, islanding detection, and secondary control of microgrids \cite{Nguyen2023, Li2023, Wang2022}. In networks dominated by inverter-based resources (IBRs), this impedance varies with topology, converter operating mode, controller action and grid strength, and it need not be equal in the three phases.

Impedance measurement (IM) methods differ in model domain, excitation and reference frame. Frequency-domain active methods inject broadband sequences \cite{Roinila2014,Roinila2018} or interharmonics \cite{Cespedes2014}. Their spectral interpretation is clear, but finite windows limit the response speed, and the injected signals may affect power quality or interact with other converters. Time-domain active methods perturb power or control references and solve the resulting equations algebraically \cite{Fang2021,Ciobotaru2007,Timbus2007,Yu2023,Mohammed2023VSG} or recursively with EKF/RLS \cite{Alenius2023,Mohammed2021,Mohammed2023,Hoffmann2014}. Passive methods rely on inherent load harmonics \cite{Liu2021} or intrinsic synchronization transients \cite{Yu2025} and require no deliberate injection.

What a passive estimator can identify is fixed by its measurements and its parameter model. Table~\ref{tab:recent_scope} summarizes four recent formulations. Intrinsic-response RLS uses local voltage/current transients \cite{Yu2025}; PQV optimization combines distinct steady operating points \cite{Guerreiro2025}; quasi-power-circle fitting uses the grid-forming (GFM) response to a drop in short-circuit ratio (SCR) \cite{Liu2026}; frequency-immune grid-following (GFL) estimation adds filter-side measurements \cite{Yu2026}. Each returns one impedance common to the three phases.

PLL-based $dq$ estimation additionally inherits the angle dynamics of the synchronization loop \cite{Harnefors2007,Wen2016}. Processing the phase quantities directly removes this dependence, as do stationary-frame and sequence-domain models. The difficulty in the phase domain is the unmeasured grid-source voltage, which can be much larger than the impedance drop that carries the parameter information. This paper addresses how to suppress that source voltage in each phase while retaining transients that separate resistance from inductance.

The contributions are as follows.
\begin{enumerate}
\item \emph{Phase-domain regressions.} A periodic difference (PD) or an SOGI-based notch difference (SOGID) cancels the periodic source voltage in each phase, and moving-window integration (MWI) replaces the current derivative by an integral and an endpoint feature. The resulting two-parameter regressions use PCC voltage and current together with a period or notch-frequency estimate; no PLL angle is required.
\item \emph{Error analysis.} A common voltage-balance model gives the source residual of each cancellation operator under frequency mismatch and bounds the parameter error by the ratio of this residual to the conditioning of the two regression features. The bound states when a recorded transient can separate resistance from inductance.
\item \emph{Phase-resolved evaluation.} Offline evaluation on two EMTP records shows identification of unequal phase impedances in a GFM record and retention of the estimate through a synchronization disturbance in a constant-impedance GFL record. The noise study shows how measurement noise triggers unnecessary updates and how a noise-referenced rule improves GFL estimation under white noise.
\end{enumerate}

Section~\ref{sec:model} states the model and assumptions, Section~\ref{sec:abc} derives the four regressions and the error bound, Section~\ref{sec:eaarls} describes the recursive estimator, Section~\ref{sec:cases} reports the case studies, noise sensitivity and scope, and Section~\ref{sec:conclusion} concludes.

\begin{table*}[!t]\centering\footnotesize\setlength{\tabcolsep}{4pt}
\caption{Passive impedance formulations: measurements, parameter models and operating requirements.}
\label{tab:recent_scope}
\renewcommand{\arraystretch}{1.2}
\begin{tabular*}{\textwidth}{@{\extracolsep{\fill}}>{\raggedright\arraybackslash}p{.17\textwidth}>{\raggedright\arraybackslash}p{.36\textwidth}>{\raggedright\arraybackslash}p{.40\textwidth}@{}}\toprule
Method & Measurements and parameter model & Conditions for identification\\\midrule
Intrinsic-response RLS \cite{Yu2025} & PCC voltage/current and frequency tracking; common $R,L$. & Informative intrinsic transients; an extended model is needed for independent phase parameters.\\\addlinespace
PQV optimization \cite{Guerreiro2025} & Five $P,Q,V_{\rm rms}$ points; common $R,X$ and source magnitude. & Distinct steady operating points and specified feasible ranges of SCR and $X/R$.\\\addlinespace
Quasi-power circle \cite{Liu2026} & $P,Q,V_{\rm rms}$; common $R,X$. & Regulated PCC voltage and an informative GFM power trajectory during an SCR decrease.\\\addlinespace
Frequency-immune RLS \cite{Yu2026} & PCC voltage/current, capacitor voltage and filter-inductance information; common $R,L$. & Additional filter-side measurements and identification of the filter/grid relation.\\\addlinespace
PD/SOGID with MWI (this paper) & PCC voltage/current and period/notch frequency; separate $R_p,L_p$. & Fundamental-source suppression and informative transients within a diagonal phase model.\\\bottomrule
\end{tabular*}
\end{table*}

\section{Model and assumptions}\label{sec:model}
\begin{figure}[!t]\centering
\includegraphics[width=\columnwidth,height=.86\textheight,keepaspectratio]{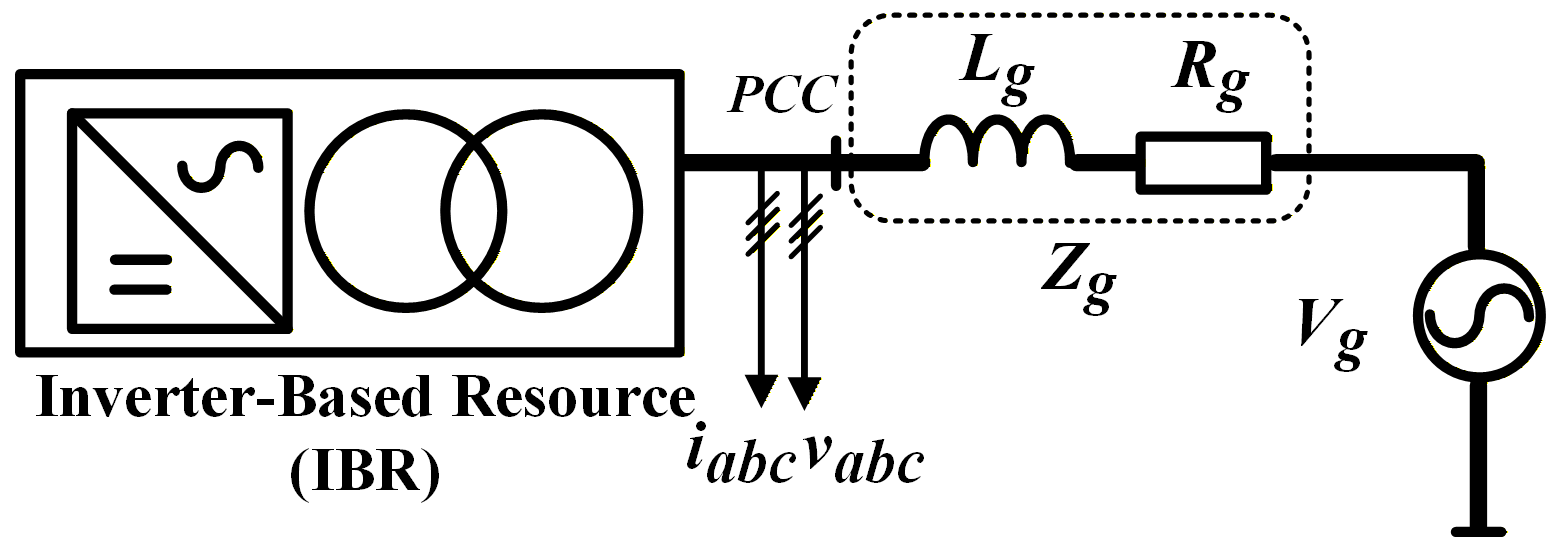}
\caption{Thevenin equivalent circuit of the grid for $R_g$-$L_g$ impedance modeling and identification.}\label{maincircuit}
\end{figure}

Fig.~\ref{maincircuit} shows the per-phase Thevenin equivalent. Positive analytical current flows from grid to PCC. Logged converter-to-grid current is converted once, so that the equivalent logged-current regression uses positive impedance-drop terms. For phase~$a$,
\par
\begin{equation}
    v_g^{a}(t) = R_g i^{a}(t) + L_g \frac{d i^{a}(t)}{dt} + v_{pcc}^{a}(t),
    \label{eq:phase_model}
\end{equation}
where $R_gi^a$ is the resistive drop and $L_g\,di^a/dt$ is the inductive drop. The source voltage $v_g^a$ is not measured and can be much larger than either drop. Phases $b$ and $c$ follow the same model with their own parameters.

The analysis rests on four assumptions.
\begin{itemize}
\item[A1)] \emph{Diagonal model.} Each phase is a series R--L branch behind a source, and the phase voltages and currents are sampled synchronously. Mutual coupling and zero-sequence paths are not represented.
\item[A2)] \emph{Periodic source.} The source voltage is periodic over the memory of the cancellation operator.
\item[A3)] \emph{Locally constant parameters.} $R_g$ and $L_g$ do not change within the operator and integration memory.
\item[A4)] \emph{Excitation.} The retained transient excites both regression features.
\end{itemize}
The regressions are evaluated offline using recorded waveforms.

Applying the same linear operator $\mathcal A$ to voltage and current gives, under A3,
\begin{equation}
\mathcal Av_{pcc}=-R_g\mathcal Ai-L_g\mathcal A\dot i+r_g,
\qquad r_g=\mathcal Av_g.\label{eq:common_residual}
\end{equation}
Matched processing of the two channels preserves the regression structure. Accuracy additionally requires the residual source voltage $r_g$ to be small relative to the retained impedance drop. Sections~\ref{sec:pd} and~\ref{sec:sogid} quantify $r_g$ for the two operators considered, and Section~\ref{sec:bound} relates it to the parameter error.

For reference, a balanced $dq$-frame benchmark is also evaluated. With the same current convention and $J=[0,-1;1,0]$, the rotating-frame equations are
\begin{align}
    v_{g,d} &= v_{pcc,d} + R_g i_d + L_g \frac{d i_d}{d t} - \omega L_g i_q, \label{eq:d_axis}\\
    v_{g,q} &= v_{pcc,q} + R_g i_q + L_g \frac{d i_q}{d t} + \omega L_g i_d, \label{eq:q_axis}
\end{align}
where $\omega$ is the angular speed of the frame. The time-step-difference moving-window integration (TSD-MWI) benchmark applies an $n$-step difference to the discretized balance, with $\omega$ frozen over the interval:
\begin{equation}
\delta_n v_{pcc,dq}=-R_g\delta_n i_{dq}
-L_g(\delta_n D^-i_{dq}+\omega J\delta_n i_{dq})
+\delta_n v_{g,dq},\label{eq:rls_d_mwi}
\end{equation}
where $\delta_n x_k=x_k-x_{k-n}$ and $D^-x_k=(x_k-x_{k-1})/T_s$. It cancels the source approximately when $\delta_n v_{g,dq}$ is small. A variable frame speed introduces additional differences of $\omega i$, and a PLL angle error can invalidate the slowly varying source approximation during synchronization transients. If both endpoints are steady and the source is unchanged in the selected frame, the derivative terms vanish and the two-point algebraic expressions are recovered. With $\Delta_{ss}x=x(k)-x(k-n)$,
\begin{equation}
    \begin{bmatrix}
    \Delta_{ss} v_{pcc,d} \\ \Delta_{ss} v_{pcc,q}
    \end{bmatrix}
    \approx
    \begin{bmatrix}
    -R_g & \hat{\omega} L_g \\
    -\hat{\omega} L_g & -R_g
    \end{bmatrix}
    \begin{bmatrix}
    \Delta_{ss} i_d \\ \Delta_{ss} i_q
    \end{bmatrix},
    \label{eq:steady_matrix}
\end{equation}
\begin{equation}
    \begin{aligned}
        R_g &= - \frac{\Delta_{ss} v_{pcc,d} \Delta_{ss} i_d + \Delta_{ss} v_{pcc,q} \Delta_{ss} i_q}{\Delta_{ss} i_d^2 + \Delta_{ss} i_q^2}, \\
        \hat{\omega} L_g &= \frac{\Delta_{ss} v_{pcc,d} \Delta_{ss} i_q - \Delta_{ss} v_{pcc,q} \Delta_{ss} i_d}{\Delta_{ss} i_d^2 + \Delta_{ss} i_q^2}.
    \end{aligned}
    \label{eq:L_analytical}
\end{equation}
These relationships connect the benchmark to conventional $PQ$-variation methods \cite{Fang2021,Ciobotaru2007,Timbus2007,Yu2023,Mohammed2023VSG}. The benchmark returns a single $R_g$--$L_g$ pair for the three phases.

\section{Phase-domain regressions}\label{sec:abc}
\subsection{Periodic difference}\label{sec:pd}
\begin{figure}[!t]\centering
\includegraphics[width=\columnwidth,height=.86\textheight,keepaspectratio]{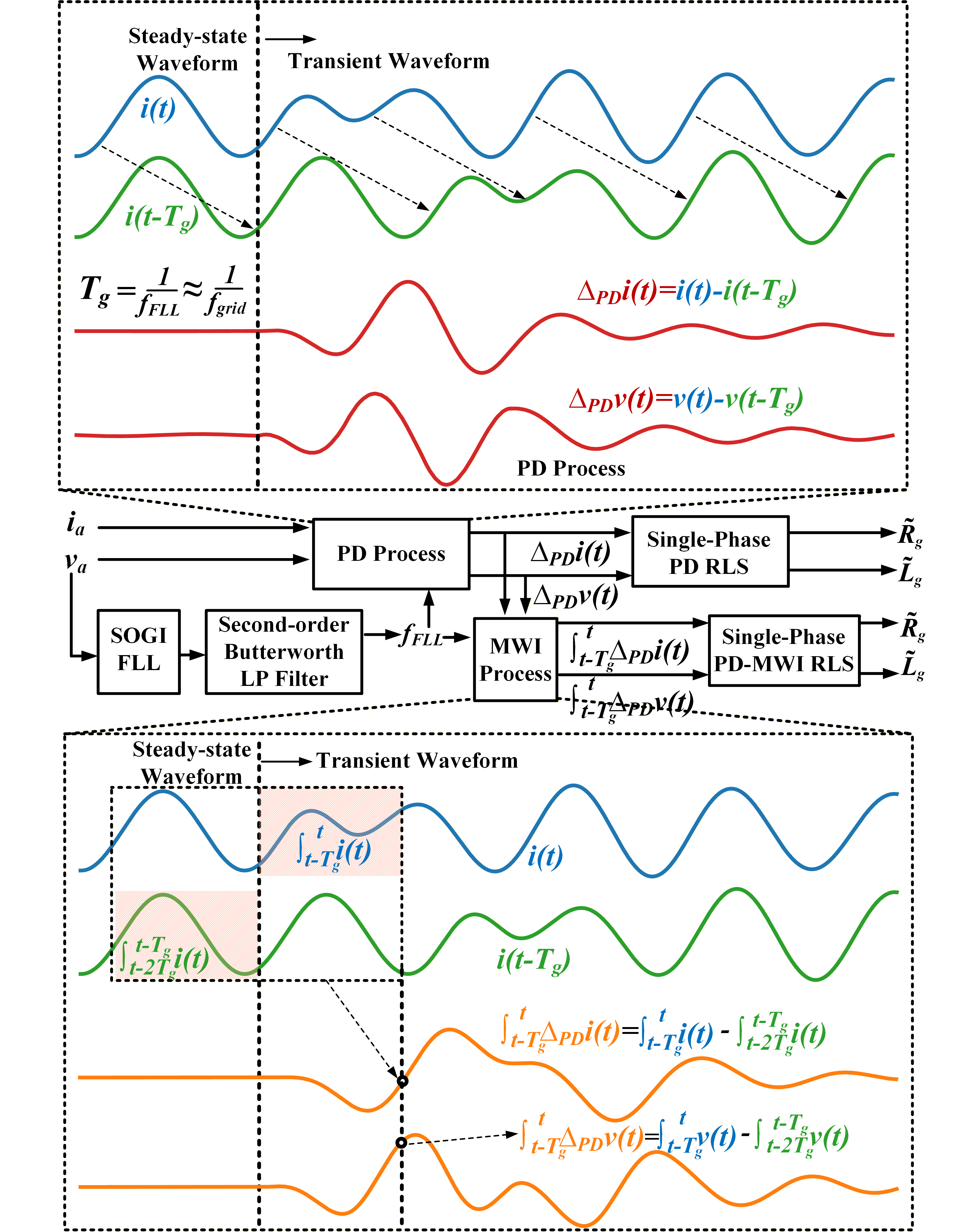}
\caption{PD and PD-MWI signal extraction. The FLL supplies the period estimate, and MWI forms integrated-voltage and current-endpoint features.}\label{PD-MWI}
\end{figure}

The PD operator cancels a quasi-periodic source by subtracting the sample taken one estimated period earlier (Fig.~\ref{PD-MWI}). Let the source be
\begin{equation}
    v_{g}^{a}(t) = E_g \sin(\omega_g t + \theta_0),
\end{equation}
and let $\hat{\omega}$ be the frequency estimated by a frequency-locked loop (FLL). With $N_p=\mathrm{round}(2\pi/(\hat{\omega} T_s))$ samples per estimated period,
\begin{equation}
    \Delta_{PD} x(k) \triangleq x(k) - x(k - N_p).
\end{equation}
From \eqref{eq:common_residual}, the PD regression is
\begin{equation}
\Delta_{PD}v_{pcc}=-R_g\Delta_{PD}i-L_g\Delta_{PD}\dot i+r_{PD},
\quad r_{PD}=\Delta_{PD}v_g.\label{eq:final_model}
\end{equation}
For a sinusoid of peak $E_g$, frequency $f$ and actual delay $\tau_d$ ($\tau_d=N_pT_s$ for an integer delay), the residual peak is
\begin{equation}
E_{r,PD}=2E_g|\sin(\pi f\tau_d)|.\label{eq:pd_error}
\end{equation}
For an ideal fractional delay $1/\hat f$, a small mismatch gives $E_{r,PD}\simeq2\pi E_g|f-\hat f|/\hat f$. At $E_g=325$~V and $\hat f=50$~Hz, mismatches of 0.01, 0.05 and 0.1~Hz leave residual peaks of 0.41, 2.04 and 4.08~V. More generally, for $v_g(t)=A(t)\sin\chi(t)$,
\begin{equation}\begin{split}
|r_{PD}|\le{}&|A(t)-A(t-\tau_d)|\\
&+2|A(t-\tau_d)||\sin(\epsilon_\chi/2)|,
\end{split}\label{eq:pd_general}\end{equation}
where $\epsilon_\chi=\chi(t)-\chi(t-\tau_d)-2\pi$. A phase-accumulating one-cycle delay bounds $|\epsilon_\chi|$ by $2\pi\tau_d\sup|f-\hat f|$, and source-amplitude steps enter through the first term.

\subsection{PD with moving-window integration}\label{sec:pdmwi}
The derivative term $\Delta_{PD}\dot i$ in \eqref{eq:final_model} amplifies measurement noise and degrades the signal-to-noise ratio of the regressor. PD-MWI removes it by integrating \eqref{eq:final_model} over a moving window. Let $\mathcal I_wx=\int_{t-T_w}^{t}x(u)du$. With a fixed delay over the window,
\begin{equation}\begin{split}
\mathcal I_w\Delta_{PD}v_{pcc}={}&-R_g\mathcal I_w\Delta_{PD}i\\
&-L_g[\Delta_{PD}i(t)-\Delta_{PD}i(t-T_w)]
+\mathcal I_wr_{PD},\label{eq:pd_mwi_integral}
\end{split}\end{equation}
so that the inductive term depends only on the current change between the window endpoints. For fixed $N_w=T_w/T_s$, backward differences pair consistently with $N_w$ right-rectangle summands. With $x_r=\Delta_{PD}x$ and $\theta=[R_g,L_g]^\top$,
\begin{equation}\begin{split}
y_k&=T_s\sum_{j=k-N_w+1}^{k}v_{r,j},\qquad
y_k=\phi_k^\top\theta+r_{w,k}+\epsilon_k,\\
\phi_k&=-\begin{bmatrix}T_s\sum_{j=k-N_w+1}^{k}i_{r,j}\\i_{r,k}-i_{r,k-N_w}\end{bmatrix}.
\label{eq:pd_mwi_final}
\end{split}\end{equation}
The first feature is the accumulated current and the second is the net current change. The term $r_{w,k}$ is the integrated source leakage, and $\epsilon_k$ collects numerical and measurement error; both remain in the integrated model. For a smoothly varying delay, $d(\Delta_{PD}i)/dt=\Delta_{PD}\dot i+\dot\tau_d\,\dot i(t-\tau_d)$ supplies the correction to the endpoint feature. The $N_w$ right-rectangle summands span the same interval $N_wT_s$ as the endpoint difference, so the discrete integral and the backward difference are consistent.

\subsection{SOGI-based difference}\label{sec:sogid}
\begin{figure}[!t]\centering
\includegraphics[width=\columnwidth,height=.86\textheight,keepaspectratio]{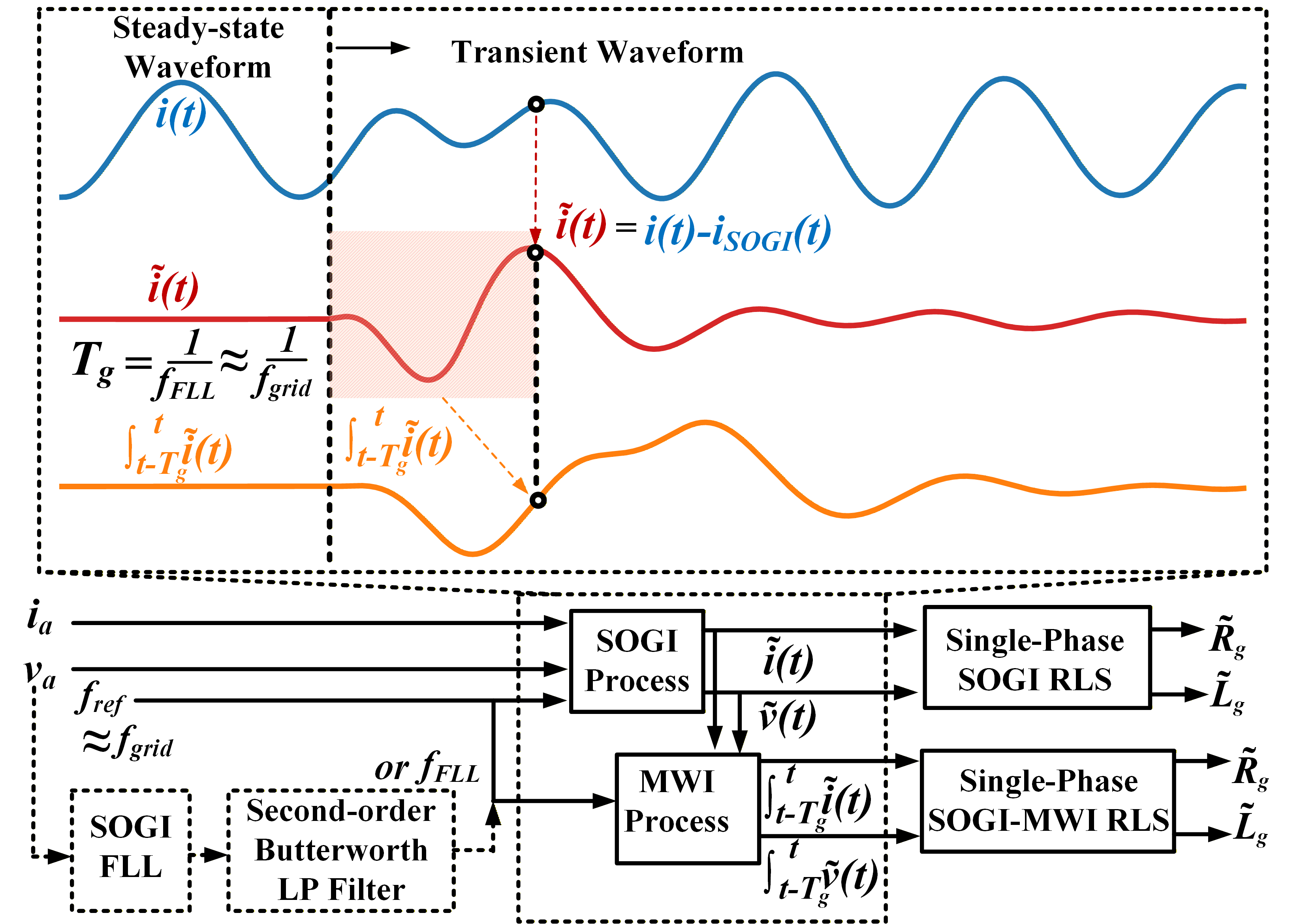}
\caption{SOGID and SOGID-MWI processing. Matched notch filters remove the nominal fundamental component before instantaneous or integrated regression.}\label{SOGI-MWI}
\end{figure}

SOGID subtracts the in-phase SOGI output from its input (Fig.~\ref{SOGI-MWI}). For a fixed center frequency $\hat\omega$ and gain $k_s$, this is the notch
\begin{equation}
H_N(s)=\frac{s^2+\hat\omega^2}{s^2+k_s\hat\omega s+\hat\omega^2}.
\label{eq:sogi_notch}
\end{equation}
Let $\tilde v=H_Nv_{pcc}$, $\tilde i=H_Ni$ and $r_N=H_Nv_g$. For a fixed LTI filter with compatible initialization, the derivative commutes with $H_N$ and \eqref{eq:common_residual} gives
\begin{equation}
\tilde v=-R_g\tilde i-L_g\frac{d\tilde i}{dt}+r_N,
\label{eq:sogi_residual_model}
\end{equation}
which has the same form as \eqref{eq:final_model}. Startup and time-varying coefficients add further terms. For a steady source at $\omega$,
\begin{equation}
\frac{E_{r,N}}{E_g}=\frac{|\hat\omega^2-\omega^2|}
{\sqrt{(\hat\omega^2-\omega^2)^2+(k_s\hat\omega\omega)^2}}.
\label{eq:sogi_detune}
\end{equation}
The residual vanishes at the center frequency and grows with the mismatch.

\subsection{SOGID with moving-window integration}\label{sec:sogidmwi}
Integrating \eqref{eq:sogi_residual_model} over the window gives
\begin{equation}\begin{split}
\mathcal I_w\tilde v={}&-R_g\mathcal I_w\tilde i
-L_g[\tilde i(t)-\tilde i(t-T_w)]
+\mathcal I_wr_N,\label{eq:integral_form}
\end{split}\end{equation}
and using $x_r=H_Nx$ in \eqref{eq:pd_mwi_final} yields
\begin{equation}
y_{N,k}=\phi_{N,k}^{\top}[R_g,L_g]^\top+r_{w,N,k}+\epsilon_{N,k}.
\label{eq:SOGI-MWI_final}
\end{equation}
PD-MWI and SOGID-MWI therefore share the same two-parameter structure and differ in their cancellation operators and residuals.

MWI forms two physically distinct features: accumulated current for R and endpoint current change for L. A longer window combines more samples in the first feature, while the second still uses two endpoints, so integration affects their noise differently. Overlapping windows also reuse samples; smoother estimates do not necessarily contain more independent information. The window should remain short enough for one R--L pair to describe the interval, since a window spanning an impedance change contains contributions from both parameter values.

\subsection{Error bound and identifiability}\label{sec:bound}
Let $S$ be a fixed positive diagonal scaling, $\theta_s=S\theta$, and let $U$ stack the rows $\phi_k^\top S^{-1}$ of a data window with weights $W$. For unregularized weighted least squares with exact regressors, locally constant parameters and nonsingular $U^\top WU$,
\begin{equation}\begin{split}
\hat\theta_s-\theta_s&=(U^\top WU)^{-1}U^\top Wr,\\
\|\hat\theta_s-\theta_s\|_2&\le
\frac{\|W^{1/2}r\|_2}{\sigma_{\min}(W^{1/2}U)}.
\label{eq:bias_bound}
\end{split}\end{equation}
The numerator in \eqref{eq:bias_bound} measures residual source and error energy; the denominator measures how well the data separate R from L. If accumulated current and endpoint change stay in a fixed ratio, different R--L combinations predict the same voltage. Useful samples vary this ratio, making the two columns of $U$ distinguishable. The smallest singular value measures the weaker of these two directions. With noisy regressors, errors-in-variables bias adds to the error described by the exact-regressor bound.

At steady sinusoidal operation, ideal PD or a settled notch tuned to the fundamental removes both source voltage and current. A large steady current then provides no new equation for R and L. A power-reference change can produce a transient that survives cancellation and supplies independent regression features. The estimator can use such operating transients without adding a test signal, provided that the retained features satisfy the independence condition above.

\section{Recursive estimation}\label{sec:eaarls}
Passive excitation is intermittent, so the estimator must decide when a new regression sample is used and how much past information is retained. Each regression $y_k=\phi_k^\top\theta$ is solved by RLS with forgetting factor $\lambda_k$:
\begin{equation}\begin{aligned}
K_k&=\frac{P_{k-1}\phi_k}{\lambda_k+\phi_k^\top P_{k-1}\phi_k},\\
e_k&=y_k-\phi_k^\top\hat\theta_{k-1},\\
\hat\theta_k&=\hat\theta_{k-1}+K_ke_k,\\
P_k&=\lambda_k^{-1}(P_{k-1}-K_k\phi_k^\top P_{k-1}),
\end{aligned}\label{eq:rls_update}\end{equation}
where $P$ is the inverse information matrix.

For a constant forgetting factor, a sample accepted $m$ updates earlier has relative weight $\lambda^m$. Its e-folding memory is therefore $-1/\ln\lambda$ accepted samples, or $-T_s/\ln\lambda$ seconds when every sample is used. A smaller factor lets a new transient replace older information sooner; a closed gate preserves the information already accumulated. The integration window defines each voltage balance; forgetting sets the weights of successive balances.

Excitation-aware adaptive RLS (EAARLS) adds two rules to \eqref{eq:rls_update}. An excitation test accepts an update only when a measure of the regressor ($\phi^\top P\phi$, $\|\phi\|^2$, or the squared operator output of the current and its increment) exceeds a threshold. A residual test switches $\lambda_k$ from a base value to a smaller fast value when $|e_k|$ exceeds a multiple of its running mean $\bar e$. A large test value indicates measured variation; whether that variation separates $R$ from $L$ is governed by the conditioning in \eqref{eq:bias_bound}.

Table~\ref{tab:audit} gives the case-specific thresholds, initial values and update settings. In the table, $\bar e$ is the running average of the absolute residual, computed with coefficient 0.99, and $i_r$ is the operator output for the current. PD-MWI and GFM SOGID increase a counter when excitation is detected and decrease it otherwise. GFL PD-MWI evaluates updates as the counter decreases. No covariance reset is used. For GFM PD-MWI, $d_z=0.005[\operatorname{var}(\Delta i_{PD})+\operatorname{var}(\mathcal I i_{PD})]$, calculated over the record. GFL SOGID-MWI starts from the 2--5-s median of SOGID, waits until 6~s before updating, and applies 0.02 output smoothing.

\begin{table*}[!tbp]\centering\footnotesize\setlength{\tabcolsep}{3pt}
\caption{Estimator settings for the GFL/GFM cases in Figs.~\ref{Test1} and~\ref{Test2}. GFL/GFM values are separated by a slash; parameters are listed in $[L,R]$ order.}
\label{tab:audit}
\renewcommand{\arraystretch}{1.25}
\begin{tabular*}{\textwidth}{@{\extracolsep{\fill}}>{\raggedright\arraybackslash}p{.13\textwidth}>{\raggedright\arraybackslash}p{.15\textwidth}>{\raggedright\arraybackslash}p{.24\textwidth}>{\raggedright\arraybackslash}p{.19\textwidth}>{\raggedright\arraybackslash}p{.21\textwidth}@{}}\toprule
Setting&PD&PD-MWI&SOGID&SOGID-MWI\\\midrule
Initial $[L,R]$&$[0,0]$&$[0,0]$&$[10^{-4},0.01]$&$[L_0,R_0]$ / $[10^{-3},0.01]$\\
Initial $P$&$100I$ / $10I$&$100I$&$\mathrm{diag}(100,50000)$&$10I$ / $1000I$\\
Base/fast $\lambda$&0.999/0.99&0.99 only / 0.9999,0.9&0.99999,0.9 / 0.99999,0.95&0.99999 only / 0.9998,0.5\\
Update gate&$\phi^\top P\phi>7\!\times\!10^{-5}$&GFL: count $\ge150$, on decrement, $\phi^\top P\phi>10^{-3}$; GFM: $\|\phi\|^2>d_z$&$i_r^2+(\Delta i_r)^2>$ 0.05 / 0.5&GFL: $t>6$ s, $i_r^2+(\Delta i_r)^2>0.5$; GFM: $\|\phi\|^2>0.2$\\
Residual trigger&$1.8(\bar e+10^{-9})$&$1.55/1.56\,(\bar e+10^{-9})$&$2.4(\bar e+10^{-3})$ / $1.6(\bar e+10^{-9})$&None / $\bar e+10^{-5}$\\
Confirmation&None&150 / 20&None / 10&None\\
Integration&None&One tracked period&None&1600 / 400 samples\\
Output smoothing&None&None / 0.05&None&0.02 / none\\
\bottomrule
\end{tabular*}
\end{table*}

\section{Case studies}\label{sec:cases}
Both records come from EMTP models \cite{EMTPGFLDoc,EMTPGFMDoc} with average-value converters rated 400~kVA, 0.38-kV line voltage and 0.6-kV dc voltage; the GFM control is PLL based. The records are interpolated to 20~kHz. The model PWM/internal-sampling settings are 20/20~kHz (GFL) and 4.5/22.5~kHz (GFM). All estimates in Figs.~\ref{Test1}--\ref{fig:archival_summary} are computed offline from the complete records. The period and notch frequency are supplied by per-phase SOGI-FLLs ($k=0.8$, $\gamma=30$, 49--51~Hz limits) with 1-Hz recursive and 0.5-Hz forward--backward smoothing, the notches are ordinary Tustin discretizations, and derivatives use centered differences. Table~\ref{tab:audit} lists the estimator settings.

\subsection{Case I: constant impedance under a GFL synchronization disturbance}
The system consists of two 400-kVA, 0.38-kV, two-level GFL converters connected to a 50-Hz, 0.38-kV equivalent grid through a common isolation transformer. Each converter is supplied by a 0.6-kV dc source and is implemented using the EMTP three-phase grid-connected converter model. The converters use decoupled-sequence $P$-$Q$ control with 20-kHz PWM and sampling and PLL-based synchronization. The voltage and current $v_{abc}$ and $i_{abc}$ are measured at the grid side of the transformer. The grid is weak, with SCR = 3. Before the reference changes, the converters operate in steady state, exchanging about $-60$~kW in total ($-50$~kW and $-10$~kW) at near-zero reactive power under the adopted sign convention.

Active-power reference changes excite the weak-grid dynamics. The measured total active power changes from approximately $-60$~kW to $-193$~kW at $t=20$~s and to approximately $333$~kW after the transition starting at $t=23$~s. A further reference change is applied at approximately $t=26$~s, after which the PLL can no longer maintain synchronism, and both active and reactive powers oscillate strongly. The total reactive power rises from near zero to several hundred kvar and, during the unstable interval, reaches the Mvar range.

\begin{figure*}[!t]\centering
\includegraphics[width=\textwidth,height=.86\textheight,keepaspectratio]{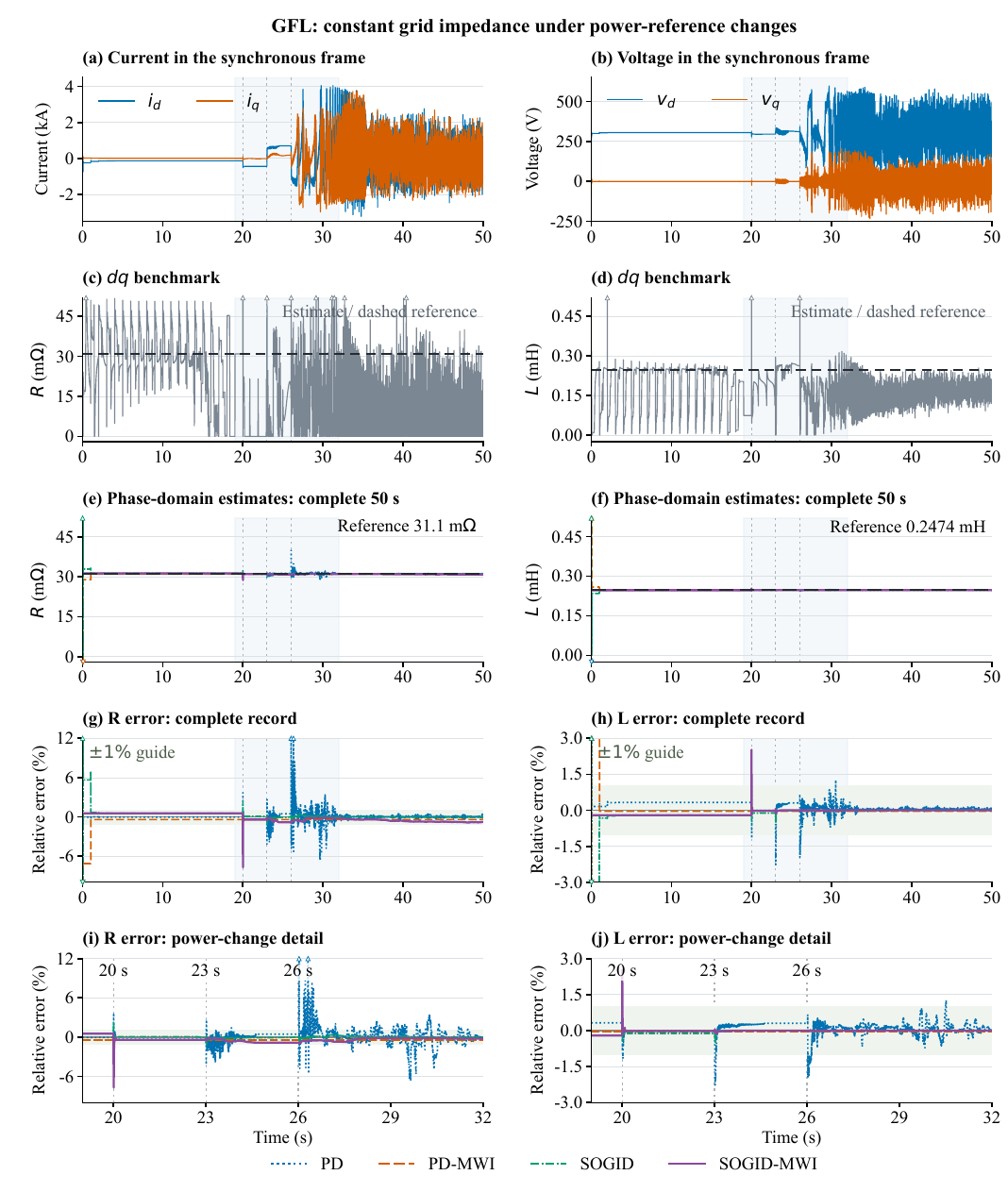}
\caption{GFL at SCR=3: rejection of power-change disturbances with constant grid impedance. (a)--(b) Synchronous-frame signals; (c)--(d) $dq$ benchmark; (e)--(f) full 50-s phase-A estimates; (g)--(h) full-record errors; (i)--(j) the shaded 19--32-s detail, with a $\pm1\%$ guide. Line styles distinguish the four methods. Triangles mark values outside the displayed range.}\label{Test1}
\end{figure*}

The grid impedance is constant throughout this record. The case therefore tests whether the estimates are retained through large disturbances; tracking of changed parameters is tested in Case~II. Fig.~\ref{Test1}(a)--(b) shows large synchronous-frame oscillations after the last power change. The $dq$ benchmark exhibits substantial excursions, whereas the phase-domain estimates remain near $R=31.1$~m$\Omega$ and $L=0.2474$~mH. PD-MWI does not update after 2~s and holds its estimate to the end of the record, with R/L errors of 0.41/0.04\%. SOGID-MWI updates during the disturbances and gives 0.56/0.06\% over 20--30~s, compared with a 1.31\% R error for PD. Over 30--50~s its errors are 0.61/0.002\%, and SOGID gives 0.0083/0.0007\%.

\subsection{Case II: varying SCR and unequal phase impedances in a GFM record}
The simulation lasts 10~s. With $Q_{\mathrm{ref}}=0$ throughout, the active-power reference $P_{\mathrm{ref}}$ (p.u.) follows the sequence $0.5,\,1.0,\,-1.0,\,1.0,\,0.5,\,1.0,\,0.5$ over the intervals $[0,1)$, $[1,2)$, $[2,3)$, $[3,5)$, $[5,7)$, $[7,9)$ and $[9,10]$~s, i.e., reference steps occur at $t=1,2,3,5,7$ and $9$~s.

The grid condition changes during the record. From 0 to 4~s the grid is balanced and strong, with SCR = 30. The SCR is reduced to 10 at 4~s and to 3 at 6~s. At $t=8$~s, with SCR = 3 maintained, phase-asymmetric grid impedances are applied, with $R_b=R_a/2$, $L_b=L_a/2$, $R_c=2R_a$ and $L_c=2L_a$. The intervals before 8~s thus probe different grid strengths under active-power excitation, and the interval after 8~s tests per-phase identification.

\begin{figure*}[!t]\centering
\includegraphics[width=.80\textwidth,height=.86\textheight,keepaspectratio]{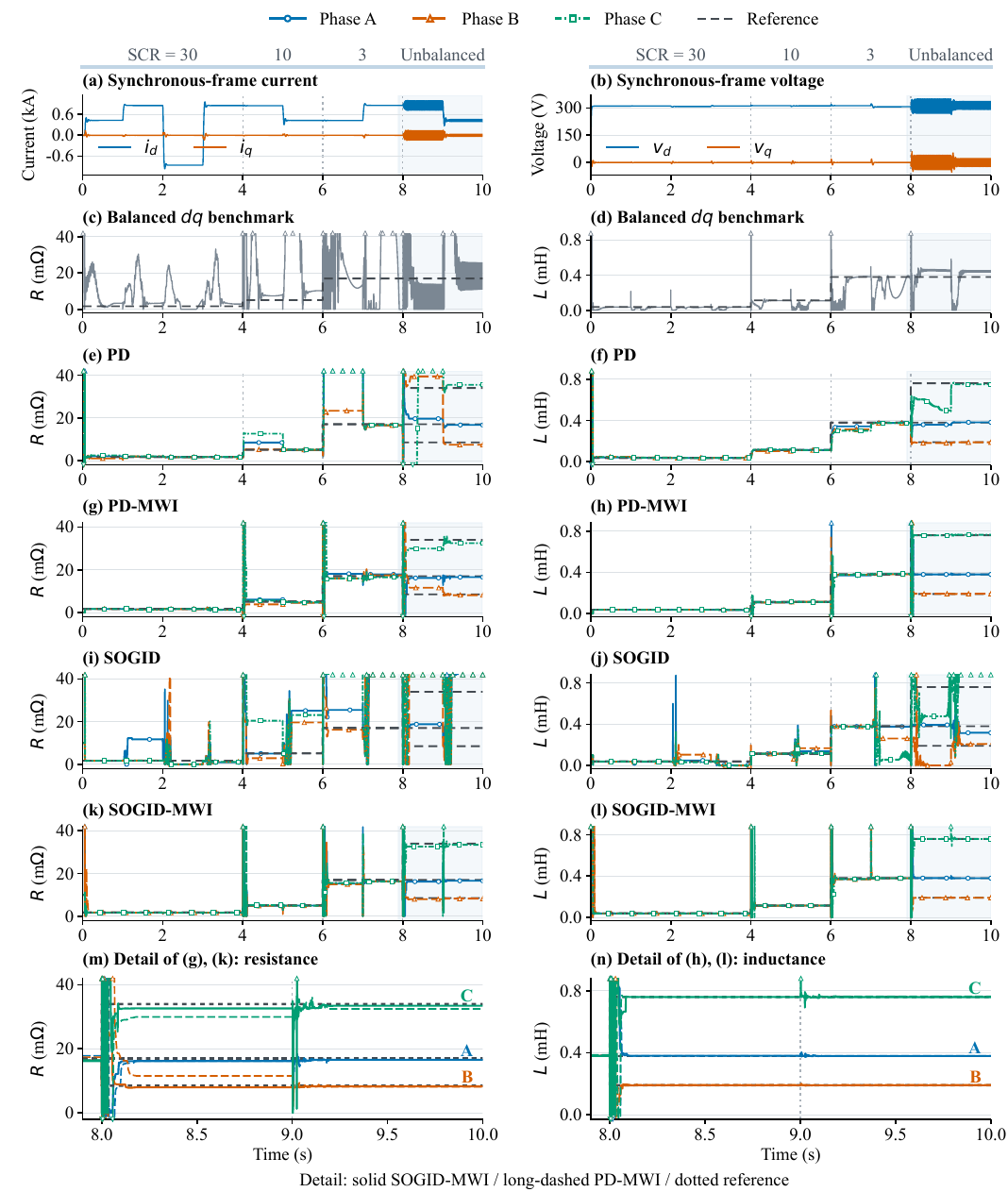}
\caption{GFM: tracking grid-strength changes and unequal phase impedances. (a)--(b) Synchronous-frame signals; (c)--(d) balanced $dq$ benchmark with phase-A reference; (e)--(l) full 10-s three-phase estimates. (m) Enlarged R estimates from PD-MWI (g) and SOGID-MWI (k); (n) corresponding L estimates from (h) and (l), both over 7.9--10~s. In the details, solid lines denote SOGID-MWI, long-dashed lines PD-MWI, and dotted black lines references; colors identify phases. Triangles mark values outside the displayed range.}\label{Test2}
\end{figure*}

\begin{figure*}[!t]\centering
\includegraphics[width=.80\textwidth,height=.86\textheight,keepaspectratio]{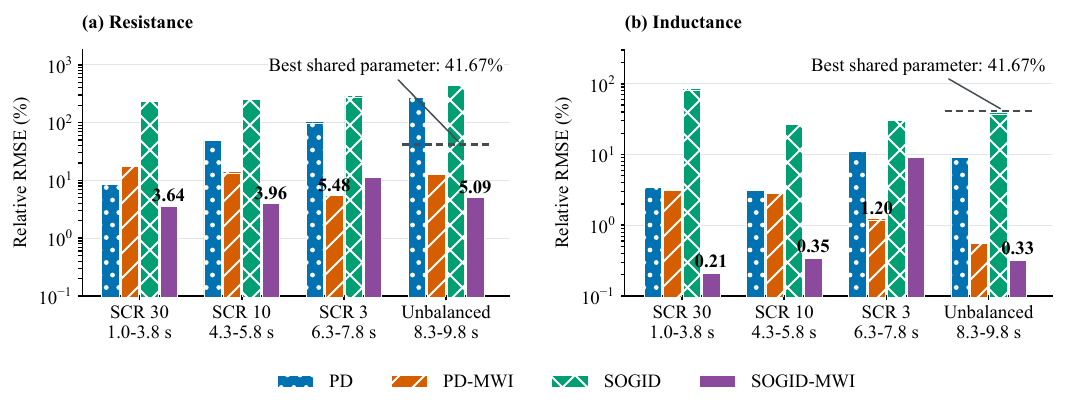}
\caption{GFM mean of phase-wise relative RMSEs. All methods and scoring intervals are shown; labels identify the lowest error. The short dashed line is the analytic minimum error of a single parameter value shared by three phases with reference ratios $1:0.5:2$.}\label{fig:archival_summary}
\end{figure*}

With the complete estimator configurations in Table~\ref{tab:audit}, Fig.~\ref{Test2} shows the responses to the grid-strength changes and the separation of the three phase values after 8~s. Fig.~\ref{fig:archival_summary} gives the arithmetic mean of the three phase-wise relative RMSEs on $[1,3.8)$, $[4.3,5.8)$, $[6.3,7.8)$ and $[8.3,9.8)$~s. These windows exclude initialization and the immediate switching transients, which remain visible in Fig.~\ref{Test2}. At balanced SCR = 3, PD-MWI gives the lower R/L errors, 5.5/1.2\%. In the unequal-phase interval, SOGID-MWI gives 5.09/0.33\%, compared with 13/0.6\% for PD-MWI.

The unequal-phase interval shows the value of separate parameters. The R and L references have ratios $1:0.5:2$. Any value shared by the three phases has a mean phase-wise relative RMSE of at least 41.67\%; for example, half the phase-A reference gives errors of 50\%, 0\% and 75\%. The dashed line in Fig.~\ref{fig:archival_summary} marks this limit, which applies to every estimator that returns one value for the three phases.

\begin{table*}[!tbp]\centering\footnotesize\setlength{\tabcolsep}{3.5pt}
\caption{Update-mechanism ablation with a shared fixed-50-Hz SOGID-MWI front end: three-phase relative RMSE (\%).}
\label{tab:noise}
\renewcommand{\arraystretch}{1.12}
\begin{tabular*}{\textwidth}{@{\extracolsep{\fill}}lllcrrrrrr@{}}\toprule
&&&&\multicolumn{2}{c}{No added noise}&\multicolumn{2}{c}{White, 40 dB}&\multicolumn{2}{c}{Correlated, 40 dB}\\
\cmidrule(lr){5-6}\cmidrule(lr){7-8}\cmidrule(l){9-10}
Record&Update&Gate&Fast $\lambda_f$&R&L&R&L&R&L\\\midrule
\multicolumn{10}{@{}l@{}}{\textit{A. Energy gating and adaptive forgetting: 10 noise realizations per record and noise type}}\\
GFM & RLS & Off & --- & 79.47 & 47.74 & 98.80 & 57.23 & 358.70 & 57.70\\
GFM & VFF & Off & 0.5 & 64.25 & 43.30 & 2113.40 & 193.76 & 2111.74 & 768.18\\
GFM & gated-RLS & Energy & --- & 57.30 & 49.95 & 96.95 & 56.52 & 325.31 & 57.17\\
GFM & EAARLS & Energy & 0.5 & 61.46 & 40.02 & 2111.74 & 190.52 & 2110.42 & 766.88\\
\addlinespace
GFL & RLS & Off & --- & 19.66 & 17.65 & 56.55 & 64.98 & 60.82 & 59.76\\
GFL & VFF & Off & 0.5 & 19.85 & 26.14 & 112.48 & 69.41 & 111.49 & 108.16\\
GFL & gated-RLS & Energy & --- & 34.08 & 73.29 & 56.02 & 64.90 & 60.25 & 59.75\\
GFL & EAARLS & Energy & 0.5 & 19.80 & 25.85 & 112.42 & 69.46 & 111.42 & 107.92\\
\midrule
\multicolumn{10}{@{}l@{}}{\textit{B. Noise-referenced updates: 2 white-noise realizations and 1 correlated-noise realization per record}}\\
GFM & EAARLS & Energy & 0.5 & 61.46 & 40.02 & 2141.25 & 187.99 & 2210.02 & 769.03\\
GFM & RLS & Off & --- & 79.47 & 47.74 & 86.27 & 57.38 & 554.51 & 57.43\\
GFM & rank-gate & Noise rank & 0.5 & 63.04 & 40.71 & 1386.16 & 165.18 & 2213.98 & 768.98\\
GFM & rank-memory & Noise rank & 0.997503 & 56.42 & 39.99 & 188.34 & 76.53 & 1648.41 & 93.31\\
\addlinespace
GFL & EAARLS & Energy & 0.5 & 19.80 & 25.85 & 112.55 & 69.45 & 111.68 & 106.85\\
GFL & RLS & Off & --- & 19.66 & 17.65 & 55.92 & 65.01 & 61.50 & 59.76\\
GFL & rank-gate & Noise rank & 0.5 & 19.43 & 25.68 & 120.50 & 69.71 & 111.73 & 107.07\\
GFL & rank-memory & Noise rank & 0.997503 & 10.16 & 0.05 & 12.27 & 11.48 & 82.85 & 57.39\\
\bottomrule\end{tabular*}
\par\smallskip\parbox{\textwidth}{\footnotesize Within each block, update rules receive identical perturbed measurements and fixed-50-Hz features. RLS is the fixed-forgetting reference. Entries average per-realization RMSEs in \eqref{eq:replay_score}; no-added-noise entries use one record. Held and bounded outputs are included. Base $\lambda=0.9998$; ``---'' denotes fixed forgetting. Rank-memory jointly changes fast memory and the residual threshold.}
\end{table*}

\begin{figure*}[!t]\centering
\includegraphics[width=.80\textwidth,height=.86\textheight,keepaspectratio]{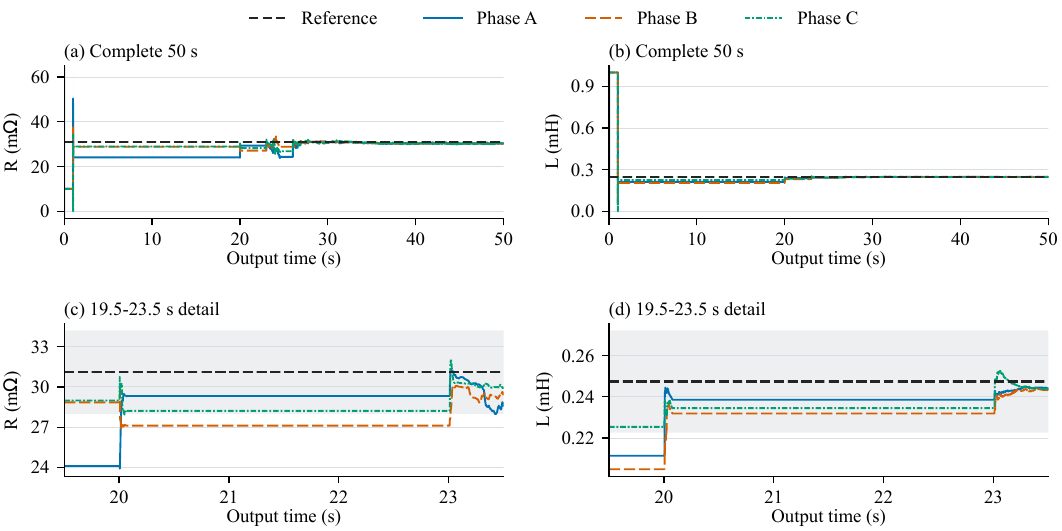}
\caption{Fixed-50-Hz rank-memory, one 40-dB white-noise realization. Top: complete 50-s GFL R/L, including initialization and holds; three phase estimates share their common reference. Bottom: same-record detail with a $\pm10\%$ reference band. The 1-ms envelopes preserve all extrema. Table~\ref{tab:noise} includes both realizations and the GFM results.}\label{fig:guard}
\end{figure*}

\subsection{Sensitivity to measurement noise}\label{sec:noise}
A controlled reconstruction tests whether measurement noise triggers updates during steady operation. All update rules share a causal prewarped 50-Hz SOGID notch ($k_s=1.414$), a 400-sample right-rectangle integration window, the same initial values and the same perturbed measurements. Independent Gaussian noise with a standard deviation of 1\% of each channel's full-record RMS (40~dB) is added to every voltage and current channel. A correlated variant uses AR(1) noise with adjacent-sample correlation 0.9 and the same nominal power. RLS is the fixed-forgetting reference; VFF changes forgetting alone, gated RLS adds an energy gate, and EAARLS combines both. The relative RMSE includes all three phases and held and bounded outputs on $[1.2,10)$~s (GFM) or $[1.2,50)$~s (GFL), where $\mathcal T$ contains $M$ scored samples:
\begingroup\setlength{\abovedisplayskip}{4pt plus 2pt minus 1pt}\setlength{\belowdisplayskip}{4pt plus 2pt minus 1pt}\setlength{\abovedisplayshortskip}{0pt plus 2pt}\setlength{\belowdisplayshortskip}{4pt plus 2pt minus 1pt}
\begin{equation}
E_j=100\sqrt{\frac1{3M}\sum_{k\in\mathcal T}\sum_{p=1}^3
\left(\frac{\hat\theta_{kpj}-\theta_{kpj}}{\theta_{kpj}}\right)^2},
\quad j\in\{R,L\}.\label{eq:replay_score}
\end{equation}
\endgroup Table~\ref{tab:noise} isolates update effects with fixed-frequency processing and whole-record scoring; Figs.~\ref{Test1}--\ref{fig:archival_summary} use frequency-adaptive offline processing and event-specific intervals. Block A of Table~\ref{tab:noise} averages ten realizations per record and noise type. Block B tests two noise-referenced variants, rank-gate and rank-memory, with two white-noise realizations and one correlated-noise realization per record. A 10-ms minimum forgetting memory and a common 201-tap 500-Hz FIR were also tested, without consistent improvement.

Three observations follow from Table~\ref{tab:noise}.

\emph{1) Noise-triggered updates.} In quiet windows of the white-noise records, EAARLS updates on 95.1\% (GFM) and 97.0\% (GFL) of the samples. Its fast factor $\lambda=0.5$ retains an e-folding memory of only 72.1~$\mu$s at $T_s=50~\mu$s, and a residual threshold near $\bar e$ is exceeded by stationary Gaussian residuals with a probability of about 0.425. As a result, EAARLS is less accurate than fixed-forgetting RLS at 40~dB in both records.

\emph{2) Noise-referenced updates.} The rank gate compares the two regression features with the noise expected after filtering and integration. Their noise levels differ, and both contain contributions from the same current samples. Accounting for this covariance allows the gate to test for independent R and L information above the noise level. The white-noise scale is estimated from the median absolute deviation (MAD) of second differences on 0.02--0.22~s, divided by $0.67449\sqrt6$. Let $g_k$ contain the integral and endpoint-difference impulse responses of the notch/MWI. The propagated current-noise covariance and the whitened feature are
\begin{equation}
C_{i,p}=\hat\sigma_{i,p}^2\sum_k g_kg_k^\top=B_pB_p^\top,
\quad z_{k,p}=B_p^{-1}\phi_{k,p}.
\end{equation}
Multiplication by $B_p^{-1}$ scales and decorrelates the features with respect to the estimated noise. For features ordered as $[R,L]$, the update condition is
\begin{equation}
q_{k,p}=\lambda_{\min}\!\left(\frac1{400}\sum_{j=k-399}^{k}z_{j,p}z_{j,p}^\top\right)>4.
\label{eq:noise_rank_gate}
\end{equation}
The minimum eigenvalue tests the less excited of the two feature directions. Requiring it to exceed 4 keeps a large signal in one direction from masking insufficient information in the other. Rank-memory also sets fast forgetting to $\exp(-T_s/0.02)=0.997503$ and requires $|e|>\bar e+4s_r$, where $s_r^2=\hat\sigma_v^2\sum_k g_{k,1}^2+\hat\theta^\top C_{i,p}\hat\theta$. Here $s_r$ estimates the residual noise standard deviation, so the residual test checks whether the mismatch exceeds the expected noise variation. All states hold when the gate closes. The 20-ms memory advances with accepted updates and pauses while the gate is closed. The memory and residual changes are tested jointly. Because successive filtered samples are correlated, the threshold is an engineering setting rather than a calibrated confidence level. On the white-noise GFL record, this rule suppresses quiet-window updates and reduces the error from 55.92/65.01\% (RLS) to 12.27/11.48\% (Fig.~\ref{fig:guard}).

\emph{3) Tracking under noise.} Neither white-noise GFM trial meets the joint $\pm10\%$ R/L criterion for 100~ms after the 4-, 6- and 8-s events. Under AR(1) noise, the variance estimate captures only 7\% of the actual variance and quiet-window gates remain open. Holds can retain biased values; voltage-only and current-only checks also point to residual source voltage and regressor noise as error sources.

\subsection{Discussion}\label{sec:scope}

\emph{Implications for impedance monitoring.} An impedance monitor must follow grid changes while retaining its estimate during converter disturbances. The GFM record tests the first requirement through changes in grid strength and unequal phase impedances; the constant-impedance GFL record tests the second through power and synchronization oscillations.

\emph{Operator characteristics.} PD removes waveforms that repeat over its one-cycle delay, whereas SOGID rejects the tracked fundamental. They therefore retain different parts of a transient. Equations~\eqref{eq:pd_error} and~\eqref{eq:sogi_detune} quantify how frequency-estimation error leaves residual source voltage in each regression and can be used to assess the required tracking accuracy.

\emph{Relation to recent passive methods.} Table~\ref{tab:recent_scope} distinguishes the required measurements and parameter models. Intrinsic-response RLS \cite{Yu2025} and quasi-power-circle estimation \cite{Liu2026} return a common impedance; the present phase model supplies three independent R/L pairs. For the available records, the GFL $X/R=2.50$ lies outside the $[3,15]$ range of PQV optimization (SCR $[1.3,10]$) \cite{Guerreiro2025}, and both records lack the filter-capacitor voltage required by \cite{Yu2026}.

The present evaluation uses a diagonal R--L equivalent at nominal frequency. Networks with mutual coupling or zero-sequence paths require additional phase parameters. Causal processing, timing tests and hardware validation are the next steps toward online use.

\section{Conclusion}\label{sec:conclusion}
This paper formulated phase-domain regressions that identify a separate resistance and inductance in each phase from recorded PCC voltage and current. PD and SOGID cancel the periodic grid source, MWI replaces the current derivative by an integral and an endpoint feature, and no PLL angle enters the regression. The voltage-balance analysis relates the parameter error to the source residual left by frequency mismatch and to the conditioning of the two features.

Over 8.3--9.8~s of the GFM record, SOGID-MWI resolved phase impedances in the ratio $1:0.5:2$ with 5.09/0.33\% R/L error, against a minimum of 41.67\% for a shared value. In the constant-impedance GFL record, the estimates were retained through a synchronization disturbance. In the fixed-50-Hz study, a noise-referenced rule reduced the 40-dB white-noise GFL error by 78/82\% relative to fixed-forgetting RLS.

\section*{CRediT authorship contribution statement}
\textbf{Songyang Zhang:} Conceptualization, Methodology, Software, Formal analysis, Investigation, Data curation, Validation, Visualization, Writing -- original draft, Writing -- review \& editing.
\textbf{Xun Gong:} Supervision, Resources, Funding acquisition, Project administration, Writing -- review \& editing.
\textbf{Bo Cao:} Supervision, Resources, Funding acquisition, Project administration, Writing -- review \& editing.
\textbf{Reza Iravani:} Supervision, Resources, Funding acquisition, Project administration, Writing -- review \& editing.

\section*{Declaration of competing interest}
Songyang Zhang is an employee of Tundra Technical Solutions Inc. and carried out this work while assigned to Huawei Technologies Canada Co., Ltd. Xun Gong and Bo Cao are employees of Huawei Technologies Canada Co., Ltd., which provided financial support for this research. Other than these disclosed employment, assignment and funding relationships, the authors declare no known competing financial interests or personal relationships that could have appeared to influence the work reported in this paper.

\section*{Funding}
This work was supported by Huawei Technologies Canada Co., Ltd. Financial support, equipment and project resources were provided within this collaboration. Co-authors affiliated with the funder participated in research supervision, manuscript review and editing, and project support.

\section*{Data availability}
The original EMTP recordings and source code used in this study are not publicly available because the commercial partner retains the rights to the data and software. The numerical results supporting the reported findings are presented in the article.

\end{document}